\documentclass[reprint,
superscriptaddress,
amsmath,amssymb,aps,showkeys,showpacs,
twoside,final,secnumarabic,%raggedbottom,
nofootinbib]{revtex4-2}

\usepackage[paperwidth=205mm,paperheight=290mm,top=17mm,bottom=25mm,
inner=17mm,outer=17mm,
twoside]{geometry}

\usepackage{cmap} % Улучшенный поиск %русских слов в~полученном pdf-файле
\usepackage[T1,T2A]{fontenc}
\usepackage[utf8]{inputenc}
\usepackage[russian,english]{babel}
\usepackage{color}
\usepackage{graphicx}% Include figure files
\usepackage{dcolumn}% Align table columns on decimal point
\usepackage{bm} % bold math
\usepackage[unicode=true,colorlinks=true,linkcolor=magenta, urlcolor=blue, citecolor = blue,breaklinks]{hyperref}
\usepackage{multirow}
\usepackage{url}
\usepackage{breakurl}
\DeclareGraphicsExtensions{.eps}

\newcount\issue
\newcount\Vol
\newcount\numb
\usepackage{fancyhdr} %this packages %provides fancy up and bottom of page
\newcommand{\ff}{\mathfrak{f}}
\newcommand{\fb}{\mathfrak{b}}

\newcommand{\cA}{\cal A}

\def\Vol{\textbf{78}}
\def\numb{x}
\begin{document}

%====== Начало шапки статьи  ============
\title{
%JOURNAL SECTION OR CONFERENCE SECTION\\[20pt]
Magnetic Catalysis for Heavy Quarks in Anisotropic Holographic Model 
%Insert your title here\\with Forced Linebreak
} 

\def\addressa{
%address 1
%\addressa
Steklov Mathematical Institute, Russian Academy of Sciences,  Gubkina str. 8, 119991, Moscow, Russia}
%\def\addressb{
%address 2
%Peoples Friendship University of Russia, Miklukho-Maklaya
 % str. 6, 117198, Moscow, Russia}

%\author{\firstname{I.Y.}~\surname{Aref'eva}}
%$\email[E-mail: ]{arefeva@mi-ras.ru}
%\affiliation{\addressa}
\author{\firstname{A.}~\surname{Hajilou}}
\email[E-mail: ]{hajilou@mi-ras.ru}
\affiliation{\addressa}
%\author{\firstname{P.}~\surname{Slepov}}
%\email[E-mail: ]{slepov@mi-ras.ru}
%\affiliation{\addressa}
 %\author{\firstname{K.}~\surname{Rannu}}
%\email[E-mail: ]{rannu-ka@rudn.ru}
%\affiliation{\addressb}

\received{January 2, 2024}
\revised{January 15, 2024}
\accepted{May 31, 2024}

\begin{abstract}
\textbf{ABSTRACT:} We consider a twice anisotropic five-dimensional holographic model
supported by Einstein-dilaton-three-Maxwell action that was constructed in  the paper \cite{ARS-Heavy-2020}. Although, that model reproduced
some essential features of the ``heavy quarks'', but did not describe the magnetic catalysis (MC) phenomenon
expected from lattice results for the Quark-Gluon Plasma (QGP) with heavy  quarks. 
We study MC phenomenon
as well as typical properties of the heavy quarks phase diagram contains magnetic field as a new parameter by  improving the holographic model, i.e. modifying the ``heavy quarks'' warp factor and the coupling function for the Maxwell field. Considering spatial anisotropy decreases the transition temperature for all values of the magnetic field for heavy quarks model.
\end{abstract}

%\pacs{Suggested PACS}\par

\keywords{AdS/QCD, phase transition, heavy quarks, magnetic catalysis, anisotropy  \\[5pt]
\textbf{DOI: 10.3103/S0027134924701455}
}
\maketitle
\thispagestyle{fancy}

%====== Начало  статьи  ============

\section{Introduction}\label{intro}
Quantum chromodynamics (QCD) is a theory that describes strong
interactions between subatomic particles such as quarks and gluons. One of the challenging questions in high energy physics is to understand the phase diagram of QCD in parameter space, i.e. temperature, chemical potential as well as magnetic field. 
In particular, to investigate the strongly coupled regime of QCD the perturbation theory does not work, while lattice theory can obtain some information but has problems with non-zero chemical potential. 
Therefore, we need non-perturbative approach \cite{Casalderrey-Solana:2011dxg,
  Arefeva:2014kyw} to study the strongly coupled quark–gluon plasma (QGP) produced in heavy ion collisions
(HIC).
%at RHIC and at the LHC.%\\

Experimental results with relativistic HIC, show that a very strong magnetic field, $eB \sim 0.3$ GeV$^2$,
is created at the early stages of the collision \cite{Bzdak:2011yy}. 
Therefore, magnetic field is a new parameter that can be considered to investigate the QCD phase diagram. Studying the QCD in the anisotropic magnetized background has received much attention recent years, because of such an interesting phenomena as chiral magnetic effect \cite{Fukushima:2008xe}, magnetic catalysis (MC) \cite{Miransky:2002rp}. Besides, some investigations that have been considered via lattice calculations \cite{DElia:2018xwo}, many
holographic models have been developed to investigate the QCD  \cite{Gursoy:2017wzz,Arefeva:2018hyo,Hajilou:2021wmz} and the effect of magnetic field on the QCD phase diagram \cite{Bohra:2019ebj,Rannu:2022fxw}. Second anisotropy, i.e. primary anisotropy produces because of non-central 
 HIC and our choice for anisopropy parameter is $\nu=4.5$ that can reproduce the energy dependence of total
multiplicity\cite{Arefeva:2014vjl}.

The MC phenomenon is the enhancement of the phase transition temperature by
increasing of the magnetic field and the
opposite effect is called inverse magnetic catalysis (IMC). 
In this research we construct a twice anisotropic ``heavy quarks'' model by considering 5-dim Einstein-Maxwell-dilaton action with three
Maxwell fields. The first Maxwell field provides finite non-zero
chemical potential in the gauge theory, the second Maxwell field
sets up the primary spatial anisotropy, and the third Maxwell field provides another
anisotropy that originates from magnetic field in the gauge theory.

\section{\label{sec:level1}
Preliminary %\textbackslash\textbackslash
}

We take the Lagrangian in Einstein frame utilized in
\cite{ARS-Heavy-2020}:
\begin{eqnarray}
  {\cal L} = \sqrt{-g} \bigg [ &R& 
    - \cfrac{\ff_0(\varphi)}{4} \, F_0^2 
    - \cfrac{\ff_1(\varphi)}{4} \, F_1^2
    - \cfrac{\ff_3(\varphi)}{4} \, F_3^2 \nonumber\\
   & & - \cfrac{1}{2} \, \partial_{\mu} \varphi \, \partial^{\mu} \varphi
    - \cal V (\varphi) \bigg ], \label{eq:2.01}
\end{eqnarray}
where $R$ is Ricci scalar, $\ff_0(\phi)$, $\ff_1(\phi)$ and $\ff_3(\phi)$ are the coupling functions associated with stress tensors of  first $F_0$, second $F_1$ and third Maxwell field $F_3$, $\varphi$ is the scalar (dilaton) field, and $V(\varphi)$ is the potential of dilaton field. 

We consider the metric ansatz as:
\begin{eqnarray}
  ds^2 = \frac{L^2}{z^2} \ \fb(z) \bigg[
    &-& \, g(z) \, dt^2 + dx_1^2 
    + \left( \cfrac{z}{L} \right)^{2-\frac{2}{\nu}} dx_2^2 \nonumber\\
    &+& e^{c_B z^2} \left( \cfrac{z}{L} \right)^{2-\frac{2}{\nu}} dx_3^2
    + \cfrac{dz^2}{g(z)} \bigg]\, \! , \label{eq:2.04} 
  \end{eqnarray}
where, $\fb (z) = e^{2{\cA}(z)}$ and for the matter fields:
\begin{gather}
  \varphi = \varphi(z), \quad \, \label{eq:2.02} \\
  \begin{split}
  F_0 : \quad   A_0 &= A_t(z), \quad 
    A_{i}=0,\,\,i= 1,2,3,4, \\
   % F_k\,- \, \mbox{magnetic ansatz} 
   F_k : \quad
    F_1 &= q_1 \, dx^2 \wedge dx^3, \quad 
    F_3 = q_3 \, dx^1 \wedge dx^2\, . 
  \end{split}\label{eq:2.03}
\end{gather}
In \eqref{eq:2.04} $L$ is the AdS-radius, $\fb(z)$ is the warp factor
is fixed by ${\cA}(z)$, $g(z)$ is the blackening function, $c_B$ is the coefficient of secondary anisotropy
related to the magnetic field $F_3$, and $\nu$ is the primary anisotropy parameter. In \eqref{eq:2.03} $q_1$ and $q_3$ are constant ``charges''.

It is very important to note that our choice for ${\cA}(z)$ determines the heavy/light quarks description of the model. For light quarks ${\cA}(z) = - \, a \, \ln (b z^2 + 1)$ \cite{Li:2017tdz}
and for heavy quarks ${\cA}(z) = - \, c z^2/4$ \cite{Arefeva:2018hyo} and $z^5$-terms \cite{Bohra:2020qom} into
the exponent warp factor. In particular, we show that $z^4$-term allows us to produce the MC phenomenon within this holographic model. 
 Supposing anisotropic metric (\ref{eq:2.04}) as an ansatz we solve Einstein equations and the field equations self-consistently by using suitable boundary condition. Finally, to solve equations of motion (EOMs) we need to fix ${\cA}(z)$ and the coupling function $\ff_0$ as:
\begin{eqnarray}
{\cA}(z) = - \, cz^2/4 \, - (p - c_B q_3)
  z^4 \\
  \ff_0 = e^{-(R_{gg}+\frac{c_B q_3}{2})z^2} \,
  \cfrac{z^{-2+\frac{2}{\nu}}}{\sqrt{\fb}} \, ,\label{eq:4.27}
\end{eqnarray}
where, we take $c = 4 R_{gg}/3$, $R_{gg} = 1.16$, $p = 0.273$, to respect the Regge spectra from lattice QCD fitting
\cite{Li:2017tdz,Yang:2015aia}.
%\\%%%%%%%%%%%%%%%%%%%%%%%%%%%%%%%%%%%%%%

%%%%%%%%%%%%%%%%%%%%%%%%%%%%%%%%%%%%%%%
\section{Numerical Results}
\begin{figure}
  \includegraphics[scale=.42]{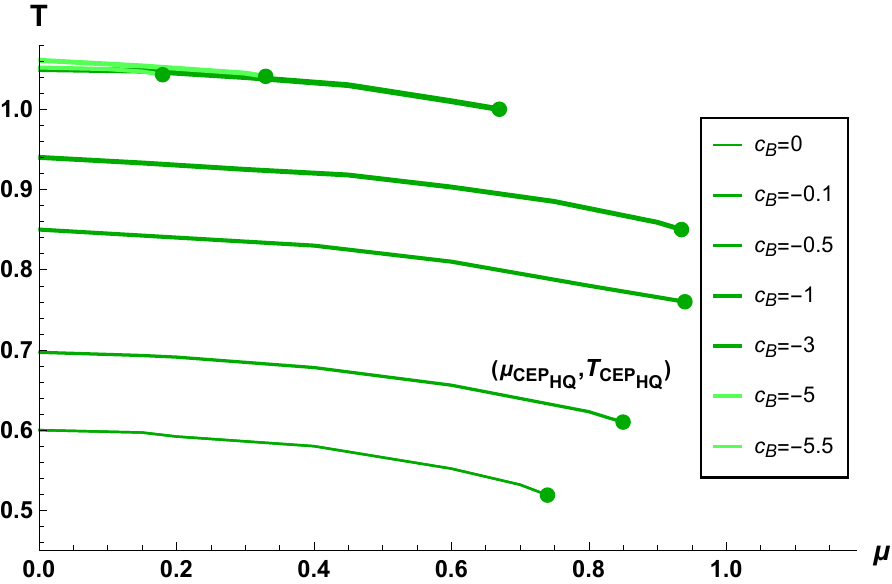} \\
 A \hspace{180pt}  \\
\includegraphics[scale=0.42]{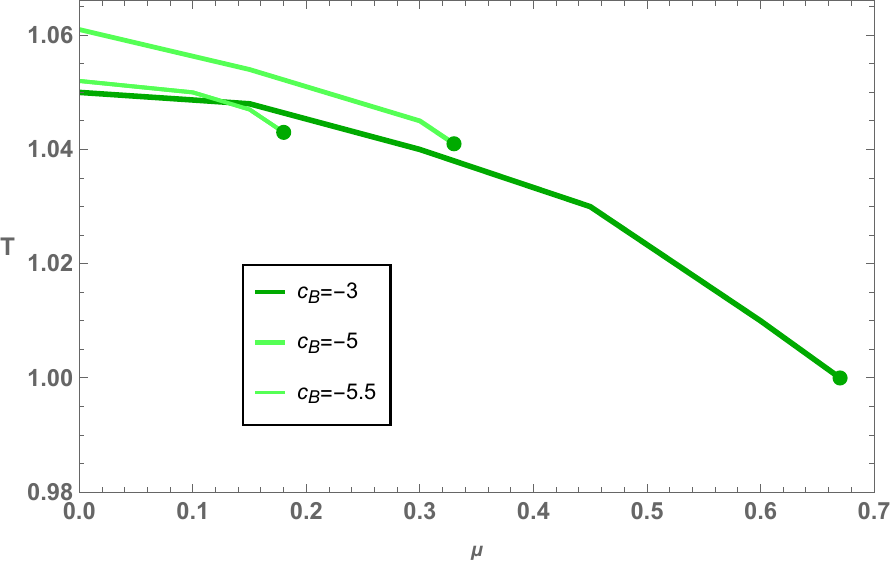}\\
 B \hspace{180pt}  \\  
\includegraphics[scale=.41]{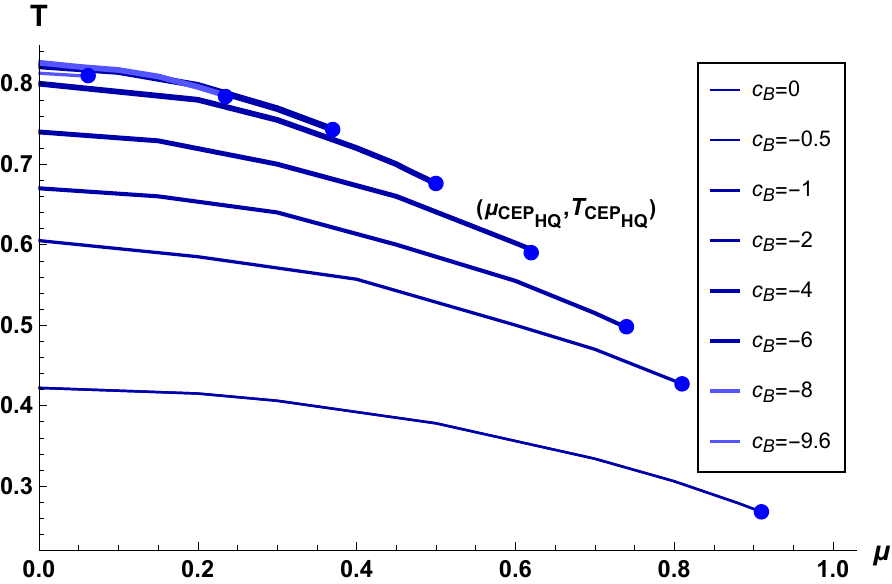}\\ 
 C \hspace{180pt}  \\
\includegraphics[scale=0.42]{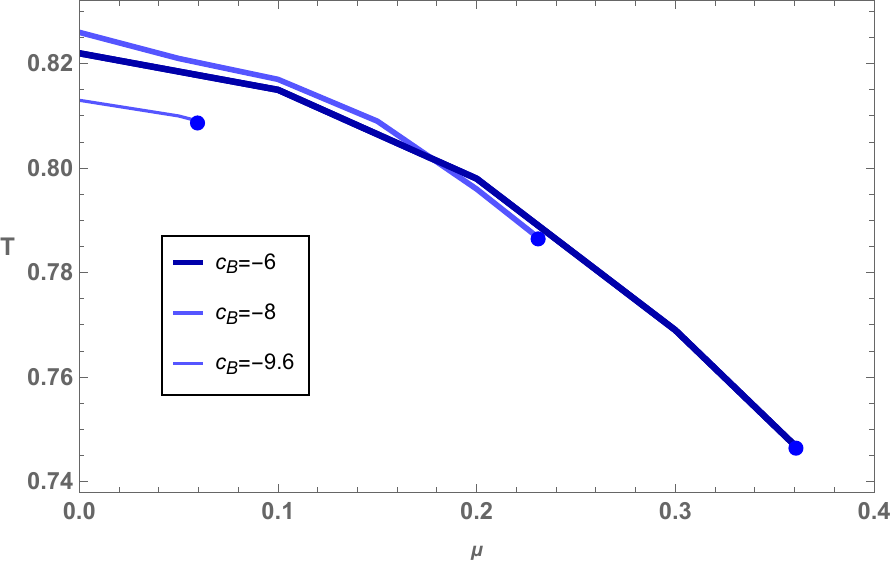}\\
 D \hspace{180pt}  \\
\caption{\label{fig:wide1}
The phase diagrams in the $(\mu,T)$-plane for heavy quarks
with different $c_B$ for $\nu = 1$ (A) and zoom in for special area of $\nu = 1$ (B), $\nu = 4.5$ (C) and zoom in for special area of $\nu = 4.5$ (D); $R_{gg} = 1.16$, $p = 0.273$, $q_3 = 5$, in units $[T] = [\mu] =$ GeV.  In all plots we considered $\nu = 1$ (green lines) and $\nu = 4.5$ (blue lines).}
\end{figure}

By solving the EOMs we can obtain the blackening function $g(z)$ and then temperature $T$ as well as free energy of the system. Then we can study the phase transition in our holographic model. In Fig.(\ref{fig:wide1}-A) panel A by increasing $|c_B|$ the critical transition temperature increases, i.e. the MC phenomenon. Interestingly, we found that at $|c_B| \sim \,6$ for isotropic case the phase transition line completely disappears and also for very small range $5<|c_B|<6$ we observe IMC phenomenon Fig.(\ref{fig:wide1}-B). The MC phenomenon also is found for primary anisotropic background Fig.(\ref{fig:wide1}-C). But, in this case at $|c_B| \sim \,10$ the phase transition line completely disappears and the IMC for the range $8<|c_B|<10$ can be observed Fig.(\ref{fig:wide1}-D).

In this research we investigated the effect of magnetic field on the first order phase transition. The end of its line $\bigl(\mu_{max},T(\mu_{max})\bigr)$  is named a critical end point ($CEP_{HQ}$ for
heavy quarks). The CEP chemical potential 
$\mu_{CEP_{HQ}}$ enhances by increasing the  $|c_B|$ for the region $0< \,|c_B| < \,0.5$ for $\nu = 1$ and after that by increasing $|c_B|$ $\mu_{CEP_{HQ}}$ decreases. For the anisotropic $\nu = 4.5$ case the $\mu_{CEP_{HQ}}$ decreases by increasing $|c_B|$.

\begin{figure*}[t!]
  \includegraphics[scale=.44]{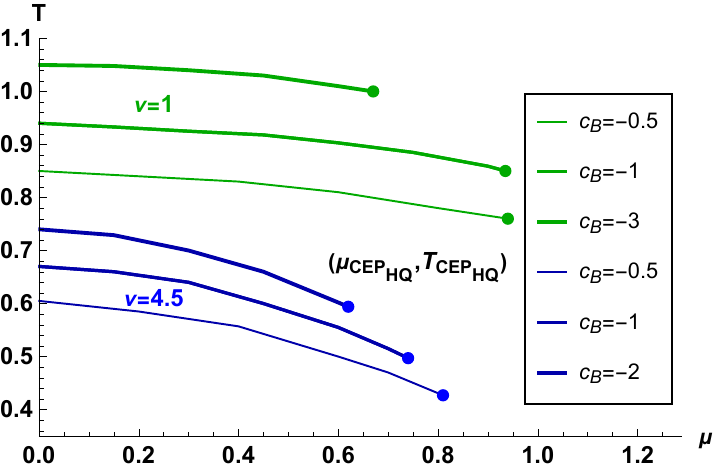} 
 \qquad 
\includegraphics[scale=0.24]{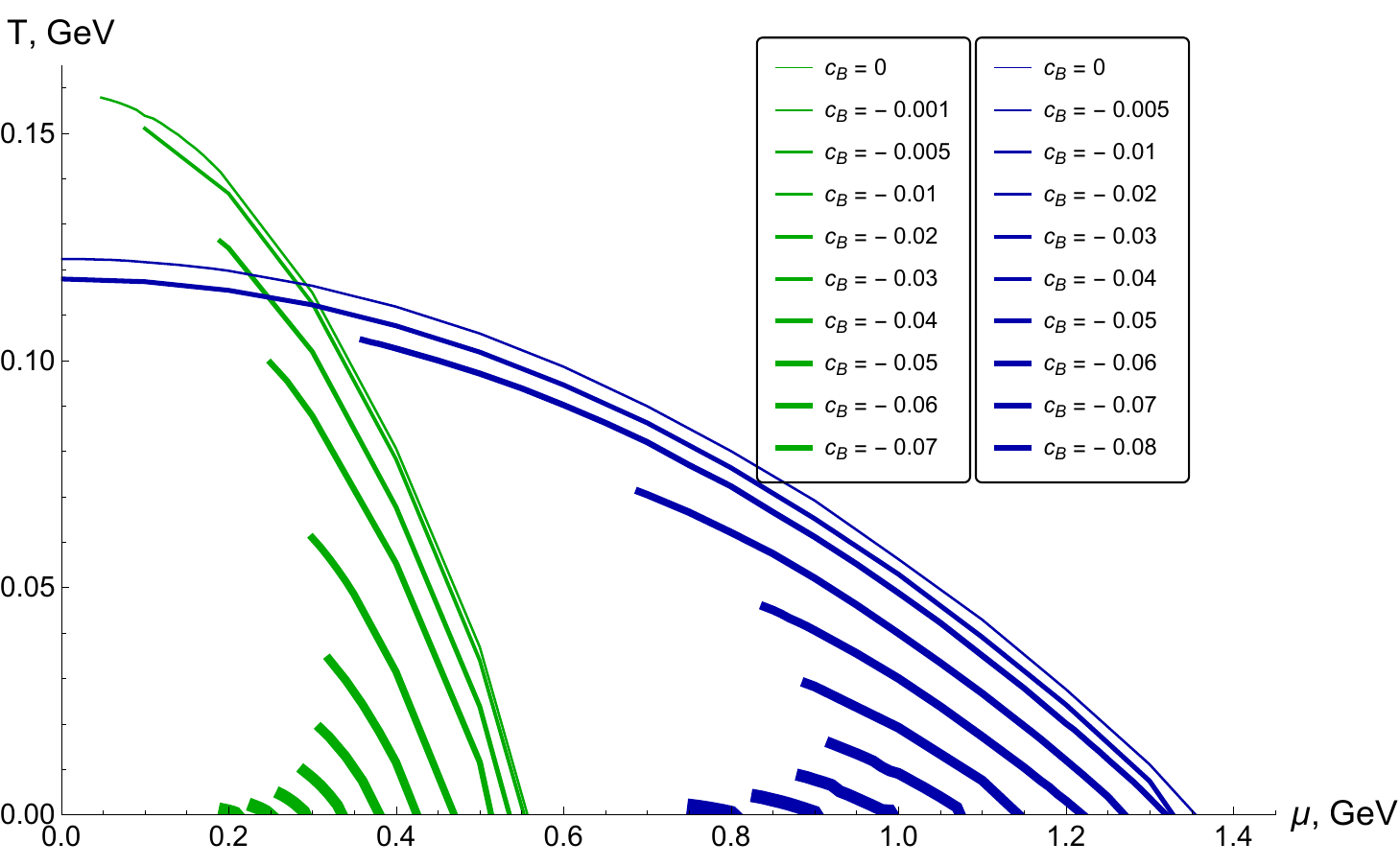}\\
  A \hspace{200pt} B \\
\caption{\label{fig:compare}
The phase diagrams in the $(\mu,T)$-plane for heavy quarks with different $c_B$. Comparison between $\nu=1$ and $\nu=4.5$ in our heavy quark model (A)
Comparison between $\nu=1$ and $\nu=4.5$ for light quark model from \cite{ARS-Light-2022} (B).  In all plots we considered $\nu = 1$ (green lines) and $\nu = 4.5$ (blue lines).}
\end{figure*}

To understand the effect of primary anisotropy on the first order transition line and phase transition temperature, we compared the $\nu=1$ and $\nu=4.5$ in Fig.(\ref{fig:compare}-A). At each fixed value of $|c_B|$ the first order transition line in isotropic case is less than anisotropic one as well as CEP. It is interesting to note that for light quarks model \cite{ARS-Light-2022} in Fig.(\ref{fig:compare}-B) we observe the IMC phenomenon in spite of MC phenomenon for heavy quarks. 
All the critical end points for $\nu=1$ and $\nu=4.5$ are mentioned in the table (\ref{table}). We showed that choosing the suitable warp factor has very crucial effect in the holographic set up to mimic the QCD characteristics via bottom-up approach.

\begin{table*} [t!]
\renewcommand{\arraystretch}{1.05}
\renewcommand{\tabcolsep}{3pt}
  \centering
  \begin{tabular}{|l|c|l|c|}
    \hline
    \,\ $\nu=1$ & $(\mu_{CEP_{HQ}}, T_{CEP_{HQ}})$ &\,\ $\nu = 4.5$ &
    $(\mu_{CEP_{HQ}}, T_{CEP_{HQ}})$ \\ \hline
    $c_B=0$      & (0.74, 0.52) & $c_B=0$      & (0.91, 0.27) \\ \hline
    $c_B=-\,0.1$ & (0.85, 0.61) & $c_B=-\,0.5$ & (0.81, 0.43) \\ \hline
    $c_B=-\,0.5$ & (0.94, 0.76) & $c_B=-\,1$   & (0.74, 0.50) \\ \hline
    $c_B=-\,1$   & (0.93, 0.85) & $c_B=-\,2$   & (0.62, 0.59) \\ \hline
    $c_B=-\,3$   & (0.67, 1.00) & $c_B=-\,4$   & (0.50, 0.70) \\ \hline
    $c_B=-\,5$   & (0.33, 1.04) & $c_B=-\,6$   & (0.36, 0.75) \\ \hline
    $c_B=-\,5.5$ & (0.18, 1.04) & $c_B=-\,8$   & (0.23, 0.79) \\ \hline
    ---          & ---          & $c_B=-\,9.6$ & (0.06, 0.81) \\ \hline
  \end{tabular} 
  \caption{The critical end points for different $c_B$ with $\nu =
    1$ and $\nu = 4.5$}\label{table}
\end{table*}

%\newpage
%\newpage

\begin{acknowledgments}
A. Hajilou wish to acknowledge I. Ya. Aref'eva, P. Slepov and K. Rannu for valuable discussions.
\end{acknowledgments}
\section*{FUNDING}
This work was performed at the Steklov International Mathematical Center and supported by the Ministry of Science and Higher Education of the Russian Federation (agreement no. 075-15-2022-265).
\section*{CONFLICT OF INTEREST}
The author declares that he has no conflicts of interest.

% The \nocite command causes all entries in a bibliography to be printed out
% whether or not they are actually referenced in the text. This is appropriate
% for the sample file to show the different styles of references, but authors
% most likely will not want to use it.
\nocite{*}

%%%%%%%%%%%%%%%%%%%%%%%%%%%%%%%%
% USE thebibliography
%%%%%%%%%%%%%%%%%%%%%%%%%%%%%%%%

\end{document}